\documentclass[a4paper]{eccomas_2022}
\usepackage{amsfonts}
\usepackage{amssymb}
\usepackage{graphicx}
\usepackage{amsmath}
\usepackage{enumitem}
\usepackage[ruled, linesnumbered]{algorithm2e}

\usepackage{tabularx}

\newcommand\clearrow{\global\let\rowmac\relax}
\usepackage[breaklinks]{hyperref}
\usepackage{caption}
\usepackage{subcaption}
\usepackage{wrapfig}
\usepackage{comment}
\usepackage{booktabs}

\title{Control of a Wind-Turbine via Machine Learning techniques}

\author{L. Schena$^1$, E. Gillyns$^1$, W. Munters$^1$, S. Buckingham$^1$, M. A. Mendez$^1$}

\heading{L. Schena, E. Gillyns, W. Munters, S. Buckingham, M. A. Mendez}

\address{$^{1}$ von Karman Institute for Fluid Dynamics\\
	Waterloose Steenweg 72, 1640 Sint-Genesius-Rode, Belgium\\
	e-mail: lorenzo.schena@vki.ac.be
}


\abstract{This article presents two \emph{model-free} controllers for wind-turbine torque and pitch control. These controllers are based on reinforcement learning (RL) and Bayesian optimization (BO) and do not rely on any mathematical model of the wind-turbine dynamics, in contrast to classical approaches designed on linearized models. The model-free controllers were benchmarked against a proportional-integral-derivative (PID) regulator in a numerical environment using Blade Element Momentum theory for computing the aerodynamic torque and the blade loads. The results showed that the model-free approaches could increase power harvesting while reducing wind turbine loads.}

\begin{document}

\section{INTRODUCTION}
\begin{figure}
\centering
\includegraphics[width=.7\linewidth]{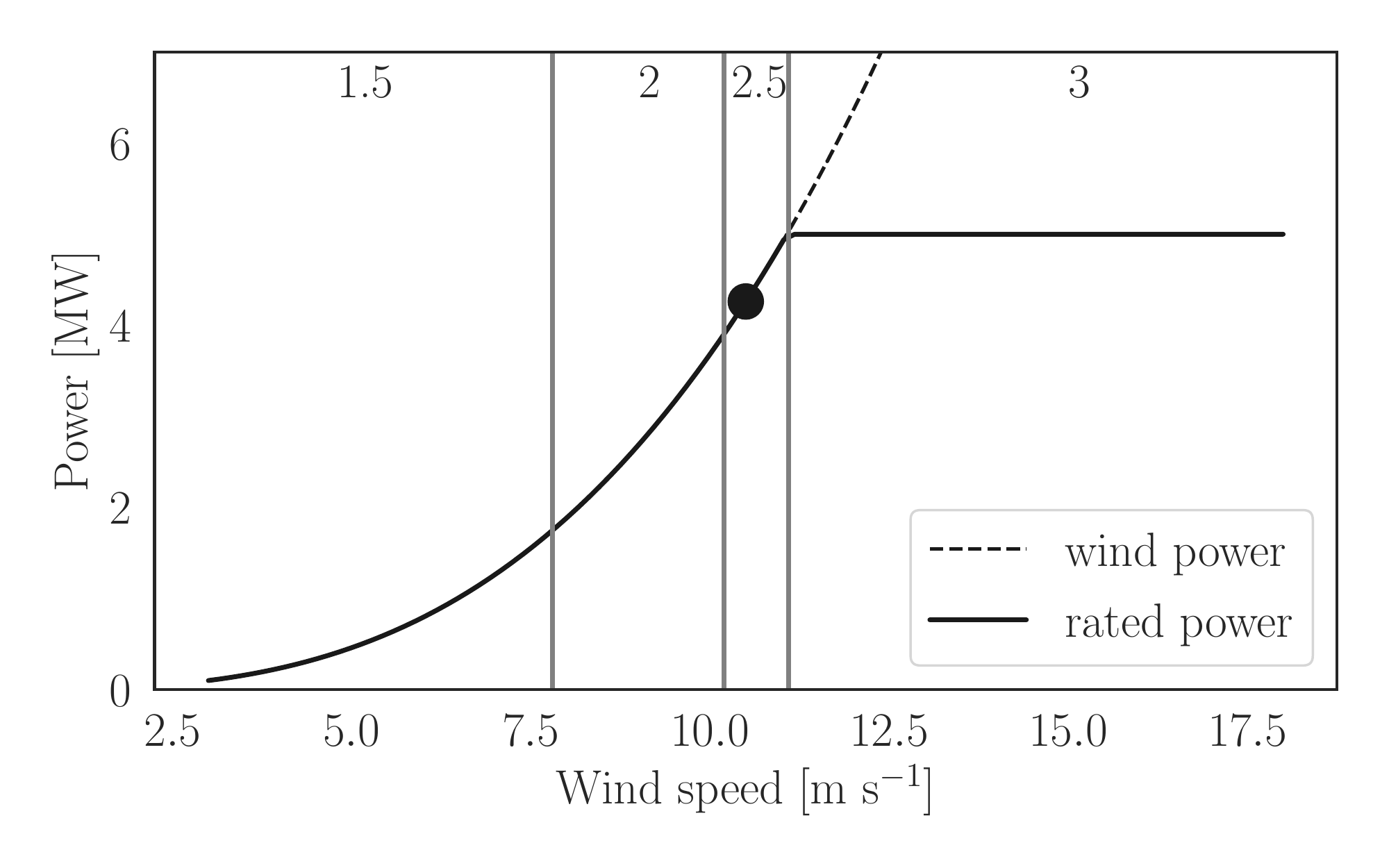}
\caption{Steady power response as a function of wind-speed for the NREL 5-MW \cite{Jonkman2009} considered in this work. The round markers indicate the investigated conditions.}
\label{fig:power-curve}
\end{figure}

The control of wind turbines is a grand challenge in wind energy \cite{Veers}, and it is crucial to its viability \cite{vanKuik}. Currently, most wind turbines operate in variable-speed or variable-pitch (VS-VP) conditions \cite{ackermann, gardner} depending on the wind speed. Figure \ref{fig:power-curve} shows the ideal power curve for a reference NREL 5-MW \cite{Jonkman2009} wind turbine considered in this study. This is designed for rated power of $P_R= 5$ [MW] at a rated wind speed of $u_R= 11.4$ [m s$^{-1}$]. A standard controller \cite{windbook} acts on the generator torque at a wind speed $u_{\infty}<u_R$, holding the blades fine-pitched to the optimal pitch angle, and on the pitch angle at $u_{\infty}>u_R$, keeping the nominal torque. This aims to maximize the power production in regions 2 and 2.5 and limit the loads in regions 2.5 and 3.
This work focuses on pitch and torque control in region 2.5, which is particularly critical in balancing the conflicting objectives of power maximization and load mitigation. 
Gusts or lulls in this region might cause undesired switches from pitch to torque control, leading to sub-optimal power production or excessive loads. Standard controllers based on PI (Proportional and Integral) formulations might offer enough flexibility in the control law to handle these conditions optimally \cite{mpc,pao1}. These controllers rely on linearized models of the wind turbine dynamics, but no linearization can handle the switch between the two operating conditions.
In this work, we investigate the application of machine learning algorithms that learn directly from data without relying on any model of the underlying dynamics, and we benchmark their performances against classical PID controllers. Section \ref{sect:methodology} overviews the simulation environment and the algorithms employed. Next, Sect. \ref{sec:4} details the test cases on which the controllers are compared. Sect. \ref{sec:5} discusses the results and Sect. \ref{sec:6} outlines the conclusions and future work.

\section{METHODOLOGY}\label{sect:methodology}
\subsection{Simulation environment}\label{subsect:sim-env}

The numerical environment considered in this work relies on a first-order model of the wind-turbine dynamics:

\begin{equation}
\label{eq:first-dof-model} 
\dot{\omega} = \frac{N_g}{J}(\tau_a - N_g\tau_g\eta_{gb})\,,
\end{equation} where $\dot{\omega}$ [rad s$^{-2}$] is the rotor acceleration, $N_g$ [-] the efficiency of the generator, $J$ is the rotor inertia [kg m$^{-2}$], $\eta_{gb}$[-] is the gearbox ratio, and $\tau_a$ and $\tau_g$ are the aerodynamic and generator torque respectively. In a first approximation, the aerodynamic torque is a function of the tip speed ratio $\lambda=\omega R/u_\infty$ and the blade pitch angle $\beta$ [deg]:
\begin{equation}
\label{eq:aero-torque}
\tau_a = \frac{1}{2} \rho \pi R^2 \frac{C_p (\lambda, \beta)}{\omega} u_\infty^3 = \frac{1}{2} \rho \pi R^2 C_Q(\lambda, \beta) u_\infty^3\,,
\end{equation}
where $C_p(\lambda, \beta)$ is the power coefficient and $C_Q = C_Q(\lambda, \beta)$ is the torque coefficient. Equation \eqref{eq:first-dof-model} and \eqref{eq:aero-torque} provide the simplest model for the wind turbine dynamics under the assumption of rigid blades and uniform wind over the rotor's swept area. Although these assumptions are rather restrictive (especially for large turbines), this model is commonly used by control engineers in wind applications \cite{Abbas2022, Johnson2004} and is thus considered appropriate for a first comparative analysis of model-free versus classic controllers.

\begin{figure}[h!]
\centering
\begin{subfigure}[b]{0.45\textwidth}
	\centering
	\includegraphics[width=\textwidth]{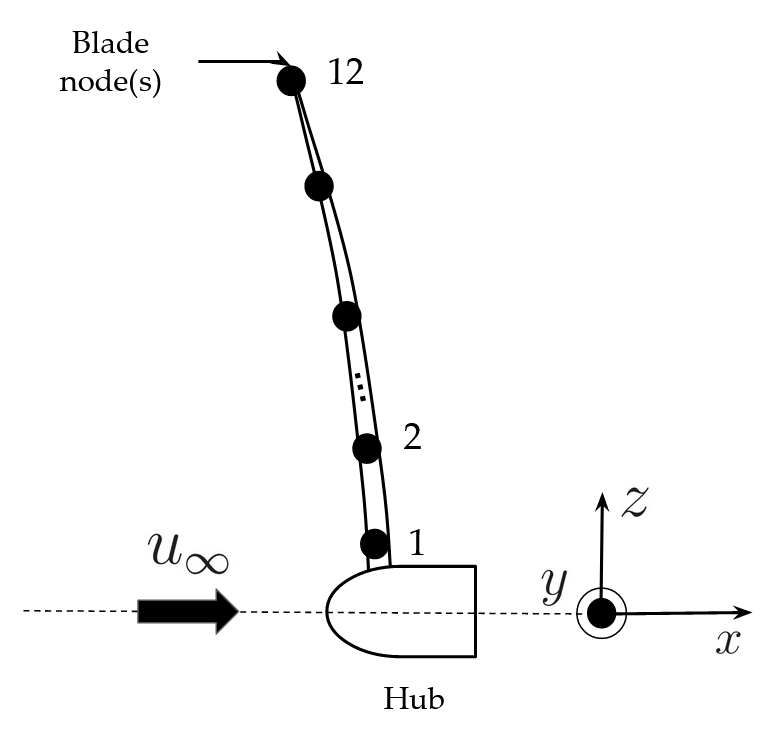}
	\caption{}
	\label{fig:hub-sri}
\end{subfigure}
\begin{subfigure}[b]{0.45\textwidth}
	\centering
	\includegraphics[width=\textwidth]{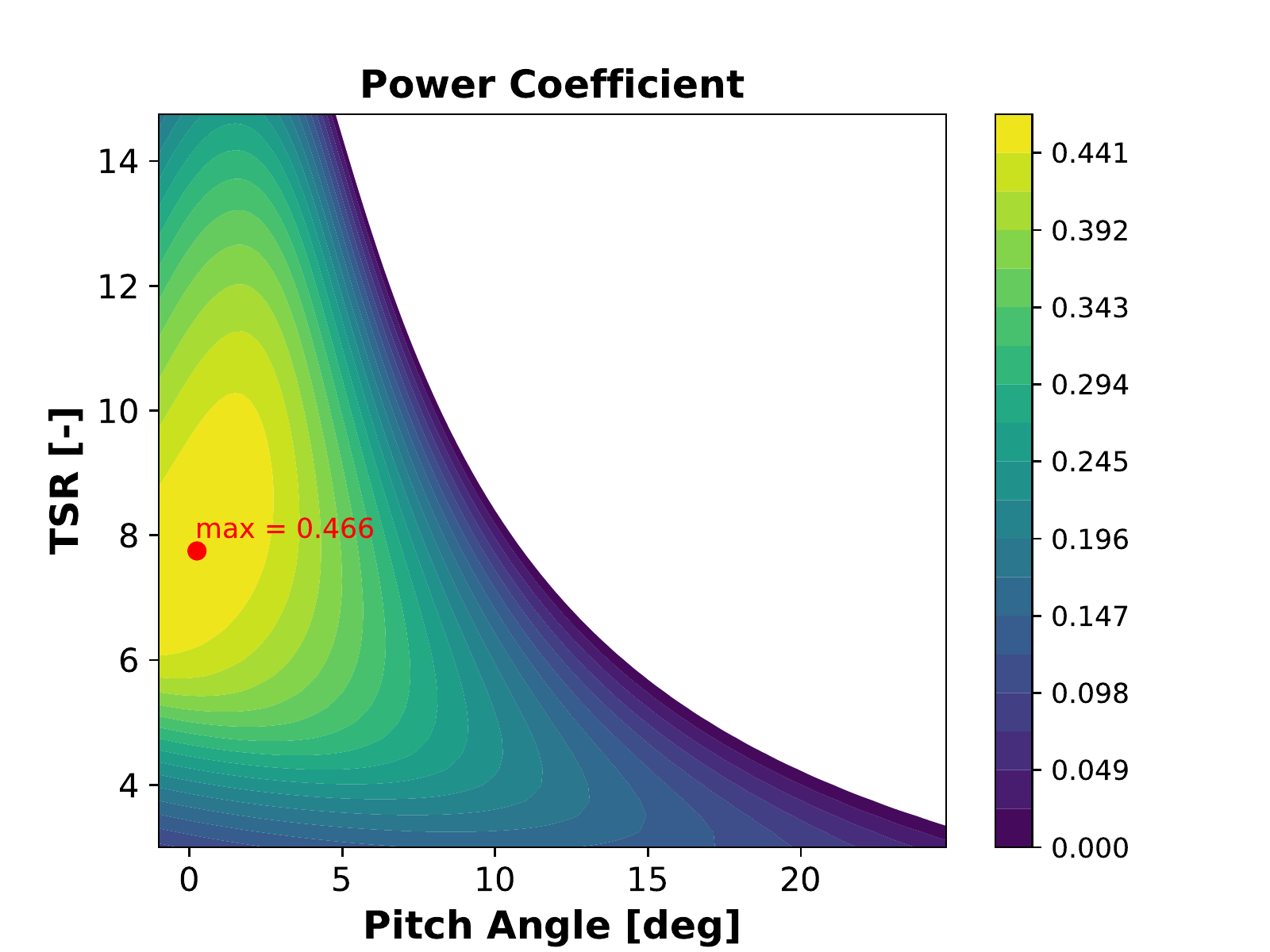}
	\caption{}
	\label{fig:cp-map}
\end{subfigure}
\caption{(a) Schematic of the blade, approximated with 12 twelve nodes and reference frame used in the BEM simulation for the test cases considered in this work. b) Contour of the resulting $C_p$ coefficient as a function of $\lambda$ and $\beta$.}
\label{fig:env-spec}
\end{figure}
In order to integrate \eqref{eq:first-dof-model} and \eqref{eq:aero-torque}, the power coefficient was computed using the Blade Element Momentum (BEM) formulation using the WISDEM CCBlade \cite{osti_1416255} \texttt{Python} package, whereas the loads have been computed using AeroDyn \cite{Jonkman2015}. This formulation couples the momentum equation with the blade element theory to compute the forces per unit length acting at different nodes (see Fig. \ref{fig:hub-sri}) along each blade. 
The wind velocity and the (instantaneous) rotational speed constitute the \emph{state} $\mathbf{s}=[u_\infty \quad \omega]^T$ vector for the dynamical system to be controlled. In order to make the system physically realistic, the controller \emph{action}/\emph{actuation} is given in terms of rate-of-change for the pitch angle ($\dot{\beta}$) and torque ($\dot{\tau}$), that is $\mathbf{a}=[\dot{\beta} \quad \dot{\tau}]^T$. Therefore, within the time integration of \eqref{eq:first-dof-model}, the update on the $\beta$ and $\tau$ under the action of the controller is:
\begin{align}
\beta (t+\Delta t)=\beta (t)+\dot{\beta} \Delta t \\  \nonumber
\tau (t+\Delta t)=\tau (t)+\dot{\tau} \Delta t
\label{eq:rate-actions}
\end{align} 
where $\Delta t$ is the time step in the integration. The rates are clipped in a range $\dot{\beta}\in[- \dot{\beta}_{max},\dot{\beta}_{max}]$ [deg s$^{-1}$] and $\dot{\tau}\in[- \dot{\tau}_{max},\dot{\tau}_{max}]$ [Nm s$^{-1}$] (see Table \ref{tab:nrel5mw-specs}).

The goal of the optimal controller is to find the \emph{policy}/\emph{actuation} law $\pi:\mathcal{S}\rightarrow\mathcal{A}$, with $\mathcal{S}\subseteq\mathbb{R}^2$ the space of admissible states and $\mathcal{A}\subseteq\mathbb{R}^2$ the set of admissible actions, such that a \emph{reward} function is maximized. The definition of this reward function depends on the control objectives and is described in Sect. \ref{sec:4}. First, we briefly introduce the investigated control strategies in Sect. \ref{sec:3}.

\begin{algorithm}[h!]
\caption{Algorithmic Approach for the Optimal Control Law Derivation}
\label{Alg_control-env}
\SetAlgoLined
\Begin{
	Initialize a possible control law $\mathbf{a}=\pi(\mathbf{s};\mathbf{w})$ of the controller\;
	Initialize the inflow velocity $u_\infty$ and add random noise, such that: $u_\infty = u_{\infty, *} + \mathcal{W}(\mu=0, \sigma)$\;
	\For {$n$ in (1, $N_{episodes}$)}{
		\For {$t$ in (1,$T$)}{
			Initialize wind-turbine at initial conditions\;
			Feed the current state $\mathbf{S} = (u_{\infty, t}, \omega_{t})$ to the controller\;
			Get action(s) $\mathbf{a} = (\beta_{t+1}, \tau_{load, t+1})$ following current policy $\pi$\;
			Evolve the wind-turbine state $\mathbf{s}_t \rightarrow \mathbf{s}_{t+1}$, evaluating the look-up-tables to get the aerodynamic coefficients $C_p, C_Q$\;
			\If{AeroDyn coupling is active}{
				Compute loads $F_x, F_y, F_z$ (in the hub reference frame) acting on the wind-turbine blades, and their rate of change $F_{xt}, F_{yt}, F_{zt}$ via BEM simulation\;
			}
			Compute the reward $r_t$ associated with the change of state $\mathbf{s}_t \rightarrow \mathbf{s}_{t+1}$ and current action $\mathbf{a}_t$\;
			Update policy weights $\mathbf{w}_k$ (for DDPG)\;
		}
		Update policy weights $\mathbf{w}_k$ (for BO)\;
	}
}
\end{algorithm}

\subsection{Algorithms}\label{sec:3}

Algorithm \ref{Alg_control-env} details the implementation of the control evaluation and training processes for the different controllers. For training purposes, the algorithm begins with random initialization of the parameters (line 2) and ends with an update of the parameters at each episode (line 14). 

The performance evaluation (whether used for training or not) is carried out within \emph{episodes} of duration $T = 100$ [s]. This also defines the duration of the simulation. This time scale is deemed adequate because it corresponds to roughly twenty blade rotations in rated conditions. Running the simulations with a time-step of $\Delta t = 0.025$ [s] and considering that the controller interacts with the system at every iteration, one episode produces 4000 agent/environment interactions per episode.
The wind-flow field is generated such that the time average is $\bar{u}_{\infty} = 10.5$ [m s$^{-1}$] (hence the turbine operates in Region 2.5, cf. Fig. \ref{fig:power-curve}), via TurbSim \cite{turbsim}. This software constructs realistic wind data from its spectral characteristic. White noise with zero mean and standard deviation $\sigma= 0.1 $[m s$^{-1}$] is also added to mimic measurement noise and avoid overfitting (line 3).
As suggested by \cite{Jonkman2009}, the velocity signal is low-pass filtered at one-quarter of the first edgewise natural frequency of the blade (corresponding to $f \approx 1$ Hz) to avoid the high-frequency excitation of these control systems. The corner frequency of the filter (that is the -3dB attenuation) is thus set to $f_c = 0.25$ Hz. The implemented filter is a 5-th order Butterworth filter with cut-off frequency $f_{co} = 1.571 / (f_s / 2)$ Hz, and $f_s = 40$ Hz is the sampling frequency.
Thus, at each timestep, the agent receives the current (filtered) state $\mathbf{s}$ and outputs an action $\mathbf{a}$ which makes the environment (plant) evolve to its next state: $\mathbf{s}_t \rightarrow \mathbf{s}_{t+1}$. An instantaneous reward signal $r(t)$ is produced at each state transition, and a cumulative reward signal $R=\sum_t r(t)$ is produced at the end of the episode. Both measure the controller performances and are maximum when the control law is optimal. The reward function definition is discussed in \ref{sec:4}. The `black-box' optimization (Section \ref{sect:bo}) updates the control law at the end of the episodes (line 17) while the reinforcement learning algorithm (Section \ref{sect:rl}) updates at each episode.
Before introducing the model-free approaches, Sect. \ref{subsect:PID} briefly recalls the fundamentals of the PID controllers used as reference.

\subsubsection{PI (or PID) controllers}\label{subsect:PID}

This is a classic control approach and serves as a reference controller in this work. The general multi-input-multi output (MIMO) formulation for the problem described in Eq. \eqref{eq:pid-law}
\begin{equation}
\label{eq:pid-law}
\mathbf{a}(\mathbf{s}) = \begin{bmatrix} k_{P1} & 0 \\ 0 & k_{P2} \end{bmatrix} \mathbf{e}(\mathbf{s}(t)) + \begin{bmatrix} k_{I1} & 0 \\ 0 & k_{I2} \end{bmatrix}\int_0^t \mathbf{e}(\mathbf{s}(\tau)) d\tau + \begin{bmatrix} k_{D1} & 0 \\ 0 & k_{D2} \end{bmatrix} \frac{d \mathbf{e}(\mathbf{s}(t))}{dt}\,,
\end{equation} where $\mathbf{e} = \mathbf{s} - \mathbf{s}_{ref}$ is the error between the current state and the reference state while the coefficients $k_P, k_I$ and $k_D$ are the proportional, integral and derivative gains respectively. The diagonal form of the matrices involved makes the controllers for each action independent from the others but one could include more general formulations with off-diagonal terms. 

Considering that in this work we do not control the first state (the wind speed), Eq. \eqref{eq:pid-law} becomes a scalar equation ($k_{P1}=k_{I1}=k_{D1}=0$). The remaining coefficients can be tuned from data or from dynamic considerations of the system at hand. In this work, we consider the second approach and we used the Reference Open Source COntroller (ROSCO, \cite{Abbas2022}) for determining $k_{P2}$ and $k_{I2}$. As it is common practice for utility-scale wind turbines, the coefficient $k_{D2}$ is set to zero to limit the control sensitivity to measurement noise. The final controller is thus a PI controller.

The reader is referred to \cite{Abbas2022} for more details on how the coefficients are derived. Briefly, these are computed from a linearized model of the wind turbine and designed to ensure that the (linear) closed loop system has a prescribed transfer function.

\subsection{Bayesian Optimization (BO)}\label{sect:bo}

The BO is a classic surrogate based approach to minimize a `black-box' function \cite{archetti2019bayesian}. As done in \cite{pinoetal}, the training of a model-free controller is converted into a black-box optimization problem by prescribing the parametric form of the control law. In this work, we consider a linear law $\mathbf{a}=\mathbf{W} \mathbf{s}$ with $\mathbf{W}\in\mathbb{R}^{2\times 2}$ collecting the four coefficients to be identified. For a given choice of parameters, the environment can be simulated for one episode and the global cumulative reward $R(\mathbf{W})$ is the cost function that needs to be optimized.

The BO formulation used in this work is the classic combination of Gaussian Process Regression (GPR) for the surrogate model and Lower Confidence Bound (LCB) for the acquisition function. The GPR was carried out using a Matern kernel with $\nu=1.5$ and length scale $l=1.0$ (see \cite{rasmussen2003gaussian} for more details).

\subsection{Reinforcement Learning (RL)}\label{sect:rl}

Reinforcement Learning (RL) is the machine learning paradigm in which the agent (controller) learns how to interact with the environment (plant) to maximize a reward signal \cite{10.5555/3312046}. The conceptual architecture of the RL agent is the same as for the BO agent: this is simply a different way of converting the problem of training a model-free controller into a black-box optimization. In modern RL approaches, based on Deep Neural Networks (also known as Deep RL), the policy $\mathbf{a}=\pi(\mathbf{s};\mathbf{w})$ is encoded in an Artificial Neural Network (ANN), and the training is performed using stochastic gradient ascent approaches. This allows for deriving a general (nonlinear) control law without the need to prescribe a specific form (like in the linear policy trained by the BO formulation). 
In this work, an in-house customized implementation of the Deep Deterministic Policy Gradient (DDPG, \cite{Lillicrap2016}) is employed (see also \cite{pinoetal}). Two ANNs are employed, one to map states to actions, the \emph{actor} and another one that approximates the long-term consequences of the applied actions - or \emph{value function}-, the \emph{critic}. The former consists in a 32$\times$32 network plus a concatenation of the input layer to allow linear control parametrizations, whereas the latter concatenates two networks. The first, from the action taken by the agent composed as 2$\times$64. The states are elaborated in two layers of size 2$\times$32$\times$32. These are concatenated and expanded by means of two layers with 32$\times$32$\times$1 , neurons, from which the output is the value estimated.
Both the network architecture and the learning approach are the same as in \cite{pinoetal}, with 2 neurons in the first (inputs) and the last (actuation) layers. The total number of parameters to be trained is 6750.

Finally, it is noted that, while the problem has been posed as a minimization problem for the BO optimization and a maximization one for the RL approach, following the relative literature practice, the two are de facto identical, as the fitness functions have been modified of sign accordingly.

\section{TEST CASES}\label{sec:4}

All the test cases investigated in this work consider the reference wind turbine NREL-5 MW with all technical details available in open-source \cite{Jonkman2009}. The relevant characteristics are reported listed in Tab. \ref{tab:nrel5mw-specs}.

\begin{table}[htbp]
\begin{tabular}{ll|r}
	\hline
	\toprule
	Rotor radius, $R$ & [m] & 63 \\
	Peak power coefficient, $C_{P, max}$ & [-] & 0.482 \\
	Tip-Speed ratio at peak power coefficient, $\lambda_*$ & [-] & 7.55 \\
	Rated mechanical power $P_r$ &[MW] & 5.296 \\ 
	Rated generator torque $\tau_r$ &[N $\cdot$ m] & 43,093.55 \\ 
	Maximum generator torque $\tau_{max}$ &[N $\cdot$ m] & 47,402.91 \\ 
	Maximum generator torque rate $\dot{\tau}_{max}$ &[N $\cdot$ m s$^{-1}$] & 15,000 \\
	Proportional gain $K_P$ at minimum blade-pitch setting, &[s] & 0.01882 \\
	Integral gain $K_I$ at minimum blade-pitch setting, &[-] & 0.0086 \\ 
	Minimum blade-pitch setting, $\beta_{min}$& [deg] & 0 \\
	Maximum blade-pitch setting, $\beta_{max}$& [deg] & 90 \\ 
	Maximum absolute blade-pitch rate, $\dot{\beta}_{max}$& [deg s$^{-1}$]& 8\\
	\bottomrule
\end{tabular}
\caption{NREL 5-MW wind-turbine main characteristics, \cite{Jonkman2009}}
\label{tab:nrel5mw-specs}
\end{table}

Two control problems are considered. The first is that of tracking the optimal Tip-speed Ratio (TSR, $\lambda$) regardless of the turbine loads. The second is that of balancing the power production with the wind blade loads. The settings for these two problems are described in Section \ref{sec:4.1} and \ref{sec:4.2} respectively.
\subsection{TSR tracking on synthetic wind data (Test Case A)}\label{sec:4.1}

This test case aims at maximizing the harvested power and the power coefficients regardless of the wind loads. 
Therefore, the instantaneous reward function is 

\begin{equation}
\label{eq:r1}
r_1(t) = - \frac{|\lambda - \lambda_*|}{\lambda_*} - \frac{|P_g - P_{*, g}|}{P_{*, g}}\,,
\end{equation} were $\lambda(t)$ is the instantaneous TSR, $\lambda_*=7.55$ is the TSR maximizing the power coefficient $C_p$ at the optimal pitch angle $\beta_*=0^\circ$ (cfr. figure \ref{fig:cp-map}), $P_g(t)$ is the instantaneous extracted power and $P_r$ is the rated power. 

It is worth recalling that the optimality of $\lambda_*$ is valid in steady conditions but not dynamic ones. In real conditions, where $C_p$ could be measured, the combination of the two terms could allow the controller to explore the action space departing from $\lambda^*$ if the associated loss in the first term is out-weighted by a gain in the second. 

\subsection{Loads mitigation (Test Case B)}\label{sec:4.2}
The goal in this test case is to minimize the overall loads met by the wind turbine.
It is recalled that to compute the loads, the augmented framework embedding BEM simulations is employed, as presented in Sect. \ref{sect:methodology}. The instantaneous reward in this second test case is 
\begin{equation}
\label{eq:r2}
r_2(t) = \langle\tilde{P}_g\rangle_T - ||\tilde{F}||_1
\end{equation}
where the $\langle\tilde{P}_g\rangle = P_{g, t} / P_{rated}$ weights the extracted power ($P_{g,t}$) on the rated power $(P_{r})$, while $||\tilde{F}||_1$ is the $l_1$ norm of the aerodynamic load in the hub reference system, as illustrated in Fig. \ref{fig:hub-sri}. The $\tilde{}$ denotes normalization of the forces with respect to the average values produced by a PI controller over an episode.

\section{RESULTS}\label{sec:5}

We analyze the performances of the model-free controllers over a training period of $40$ episodes, corresponding to $4000$ s of physical time. We here focus on the best results achieved at the end of the training and we postpone the analysis of the learning curves to the extended version of this work.

\subsection{TSR tracking on synthetic wind data}\label{sec:5.1}
The numerical results of Test case A (reward function in Eq. \eqref{eq:r1}) are listed in Tab. \ref{tab:testA}. The table collects the relative gain/increase in the power extracted, defined as $\Delta P_g= (\langle P \rangle_{PI} - \langle P \rangle_{ML})/\langle P \rangle_{PI}$, where $ \langle P \rangle_{ML}$ represents the mean power extracted in the simulation time by the control parametrization at hand, and the relative TSR tracking error, defined as $\Delta \langle \lambda \rangle =(\langle \lambda \rangle_{PI} - \langle \lambda \rangle_{ML})/\langle \lambda \rangle_{PI} $. Both quantities are averaged within an episode.

\begin{figure}[h!]
\centering
\begin{subfigure}[b]{0.45\textwidth}
	\centering
	\includegraphics[width=\textwidth]{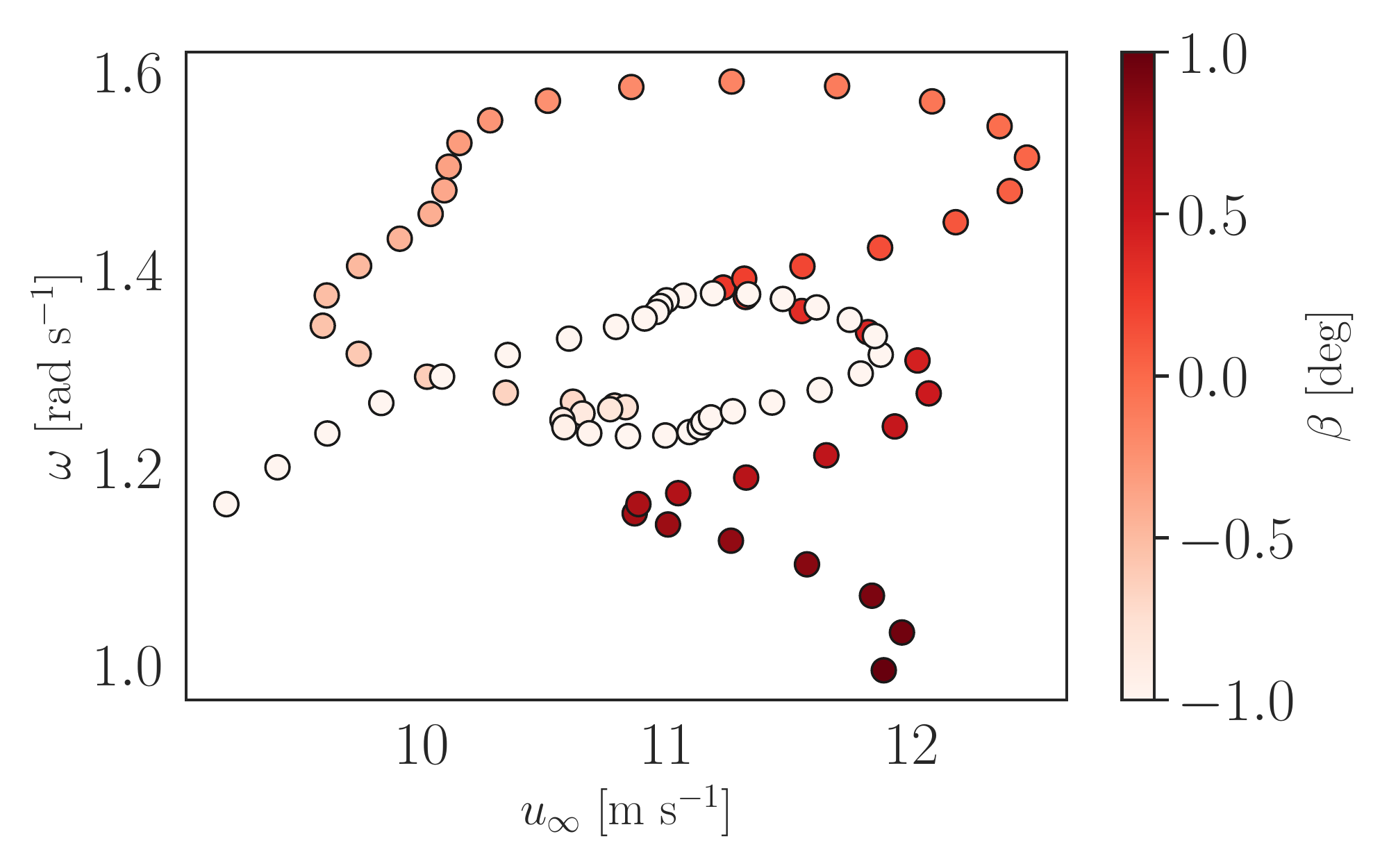}
	\caption{BO pitching policy}
	\label{fig:contour-policy-pitch-bo-105}
\end{subfigure}
\hfill
\begin{subfigure}[b]{0.45\textwidth}
	\centering
	\includegraphics[width=\textwidth]{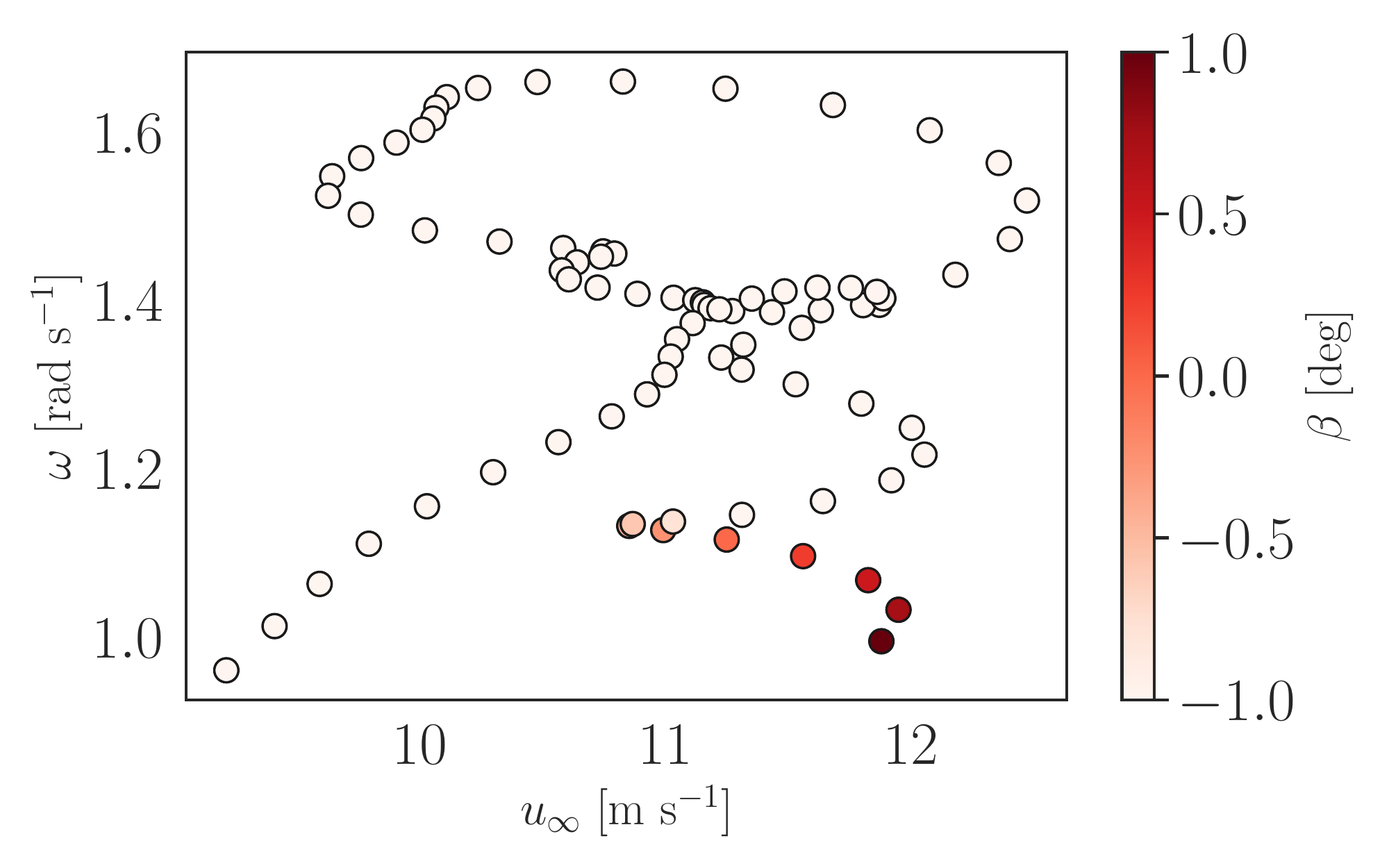}
	\caption{RL (DDPG) pitching policy}
	\label{fig:contour-policy-pitch-rl-105}
\end{subfigure}
\hfill
\begin{subfigure}[b]{0.45\textwidth}
	\centering
	\includegraphics[width=\textwidth]{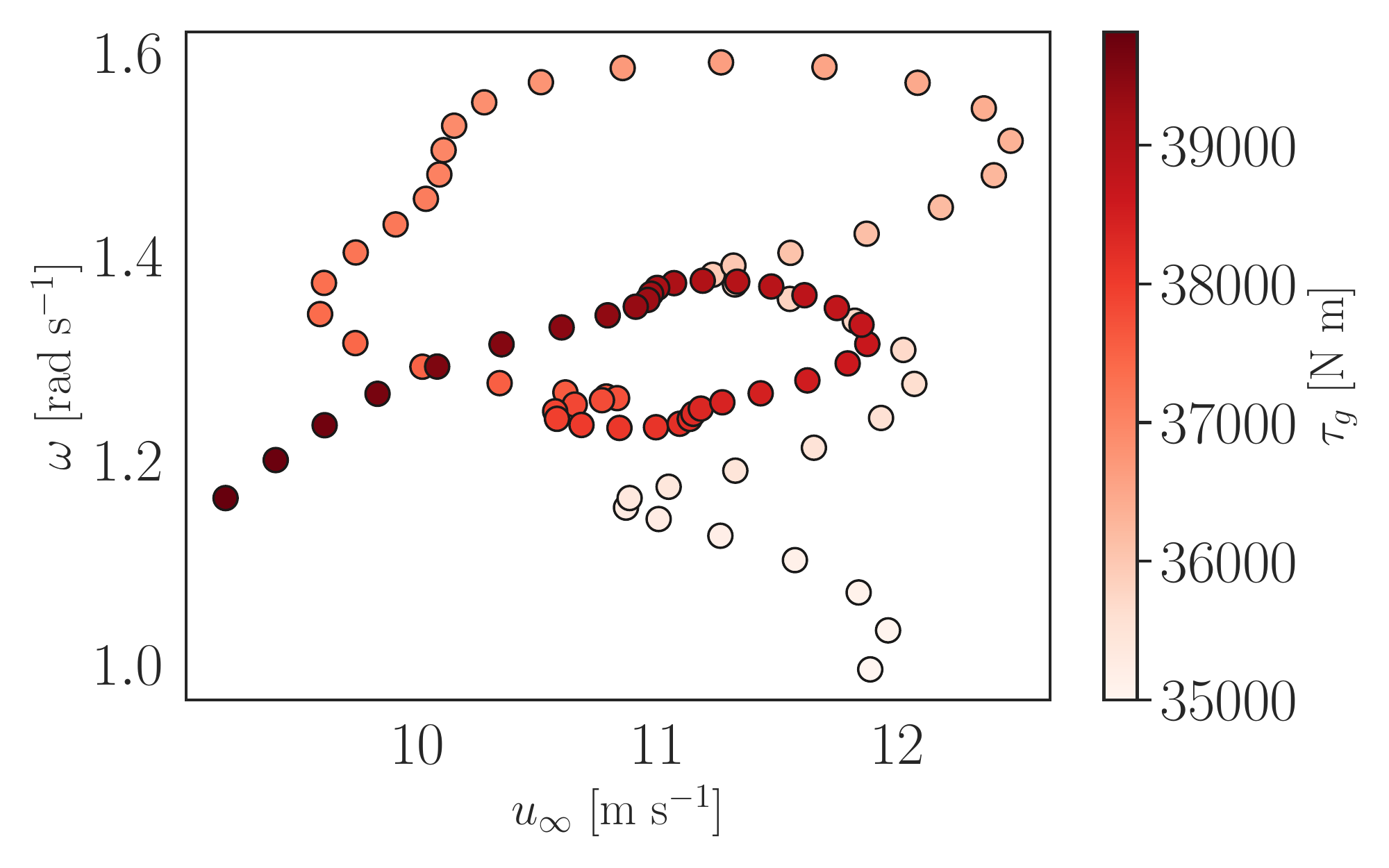}
	\caption{BO generator torque policy}
	\label{fig:contour-policy-torque-bo-105}
\end{subfigure}
\hfill 
\begin{subfigure}[b]{0.45\textwidth}
	\centering
	\includegraphics[width=\textwidth]{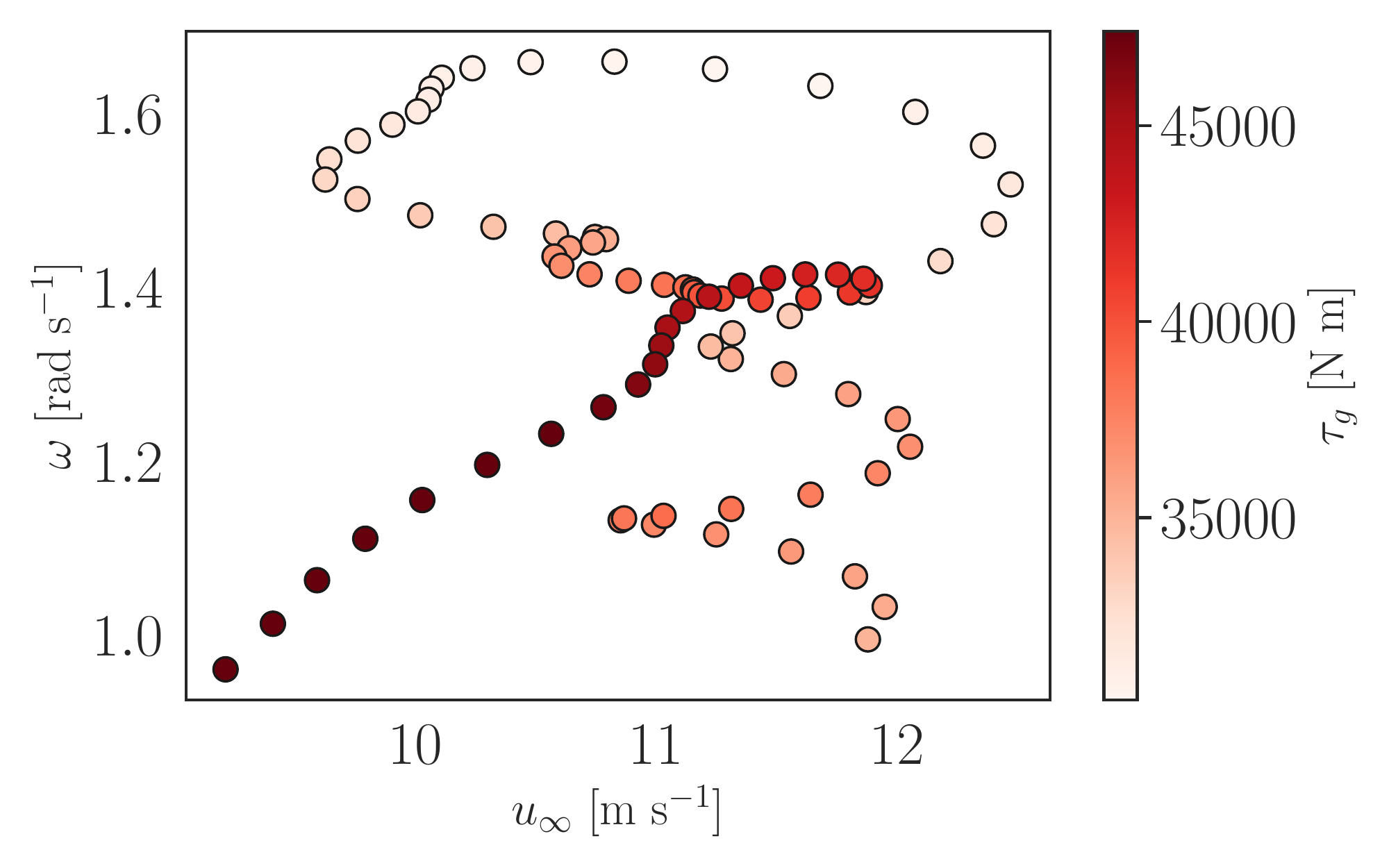}
	\caption{RL (DDPG) generator torque policy}
	\label{fig:contour-policy-torque-rl-105}
\end{subfigure}
\caption{System trajectory and controller policy for the BO (on the left) and DDPG (on the right) algorithms. The first row refers to the pitch rate actuation while the second refers to the torque rate actuation.}
\label{fig:test-A-pol}
\end{figure}

The two machine learning methods increase greatly the power extraction with respect to the PI controller while maintaining the rotational speed of the turbine closer to its rated value $\omega_r = 1.33$ rad s$^{-1}$. Interestingly, the results shows that the second term in Eq. \eqref{eq:r1} is more important than the first.
Figs. \ref{fig:contour-policy-pitch-bo-105},\ref{fig:contour-policy-pitch-rl-105}, \ref{fig:contour-policy-torque-bo-105} and \ref{fig:contour-policy-torque-rl-105} show the system trajectory in the state plane $\mathbf{s}=(u_{\infty},\omega)$ during an episode. For plotting purposes, only one every fifty time steps is included. The markers are colored by the controller actions at the corresponding states. The figures on the top show the pitch actuation ($\beta$), while the figures one the bottom show the torque actuation ($\tau$). 

\begin{table}
\centering

\begin{tabular}{llccccccccc} 
	\hline
	Method  & $\langle {P}_g \rangle$ [MW] & $\langle\lambda\rangle$ [-] & $\langle\omega\rangle$ [rad s$^{-1}$]& $\Delta P_g $\% & $\Delta \langle\lambda\rangle $\%\\ \toprule
	\midrule 
	PI & 4.05 & 7.28 & 1.22 & - & - \\
	Linear Controller (BO) 	& 4.80 & 7.61 & 1.32  & +18.1 \% & - 70.21 \% \\
	RL (DDPG) & 4.95 & 7.85 & 1.36  & + 22.22 \% & +11.1 \% \\
	\bottomrule
	
\end{tabular}
\caption{Results testcase A. Method indicates the parametrization of the control-law, $\langle P_g \rangle$ [MW] is the generated power and $\omega$ [rad s$^{-1}$] is the mean rotational speed. The last two columns compare the performances with the baseline (PID).}
\label{tab:testA}
\end{table}

Interestingly, both controllers opt for a pitch angle between $-1^\circ \le \beta \le 1^\circ$, even though they are not constrained to do so. This is the pitch angle corresponding to the highest $C_p$, for the $\lambda$ met by the operative trajectories. Concerning the BO torque actuation, the controller seeks to slow down the wind turbine when the tip speed ratio is greater than the rated value, whereas it reduces its actuation in the lower-right part of the plot, where $\lambda \approx 5.25$, to speed the rotor up. The RL acts similarly, but it surprisingly acts less in the upper part of the trajectory (corresponding to a larger tip-speed ratio than the rated value). This behaviour, which might be due to poor state space sampling, will be further investigated in future works. 
\subsection{Load mitigation}\label{sec:5.2}
The numerical results of Test case B (reward function in \eqref{eq:r2}) are listed in Tab. \ref{tab:res-loads}. Both model-free controllers significantly reduce the loads, albeit at the cost of a significant reduction in the extracted power. This result is due to the relative weights introduced in the reward function-- one could change these to set different priorities for the two objectives. By repeating the training with different relative weights between the terms in \eqref{eq:r2}, one could identify an approximation of the Pareto front in the action space \cite{pareto}.

Remarkably, the linear control driven by the BO achieves better performances in terms of load reduction for a nearly equal power reduction. Although a more in-depth analysis is needed to assess the learning performances of both agents, this result shows that the larger model capacity (and complexity) of the DDPG agent might not pay off over a more straightforward linear control law (at least within the 40 episodes considered for the training).

\begin{table*}[h!]
\centering
\begin{tabular}{llccccccc} 
	\hline 
	Method & $P_g$ [MW] & $F_x$ [kN] & $F_y$ [N] & $F_z$ [N] & $\Delta F_x$ \% & $\Delta F_{y}$ \% & $\Delta F_{z}$\\ \toprule \midrule 
	
	PI &   4.05 & 295.162 & 5.1E-02 & 3.0E-02 &- & - & - \\
	Linear Controller (BO) &	2.8 & 81.82 & 4.8E-02 & 2.1E-02 & -73\% & -7\% & -31\% \\
	RL (DDPG) & 2.78 & 142.245 & -2.8E-02 & 2.8E-02 & -53.81 \% & -44.38\% & -8\% \\ \bottomrule
\end{tabular}
\caption{Results Testcase B. Method indicates the parametrization of the control-law, $\langle P_g \rangle$ [MW] is the generated power, $F_x, F_y$ and $F_z$ are the forces met by the wind-turbine in the respective directions (in the hub frame of reference cfr. Fig. \ref{fig:hub-sri}), whereas the last three columns represent their variation against the baseline. }\label{tab:res-loads}
\end{table*}

\begin{figure}[h!]
\centering
\includegraphics[width=0.85\textwidth]{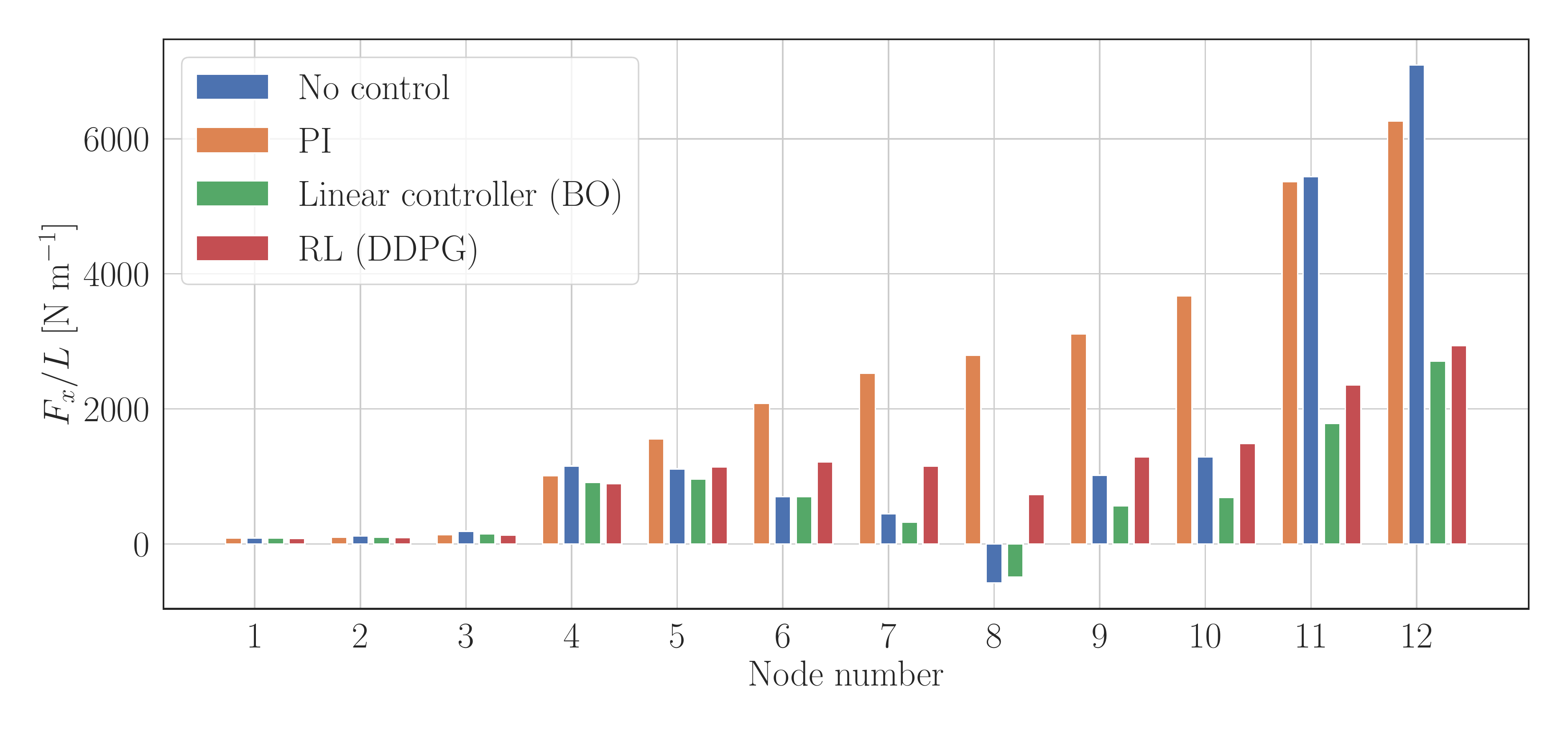}
\caption{Loads per unit length on each blade node in case of no control (blue), baseline PI control (orange), linear BO controller (green) and nonlinear (DDPG) controller}
\label{fig:loads-bar}
\end{figure}

Finally, Figure \ref{fig:loads-bar} shows the loads in each node (averaged over an episode) per unit length. The node numbering increases from the hub to the tip (cf. Figure \ref{fig:hub-sri}). The load distribution in the absence of control is shown for comparison, and it is clear that this increase from left to right (i.e. at a larger distance from the hub).
In the absence of control, the load re-partition is extremely non-uniform, with most of the loads acting on the tip while the PI controllers introduces a nearly triangular distribution.
On the other hand, the non-monotonic trend for the controlled conditions is due to a complex interplay between the time-varying accelerations produced by the controllers and the time-varying deflections of the blades. In the future, the reward function could be crafted to prescribe a specific load distribution and thus tailor the loadings to the structural design of the blades.

\section{CONCLUSIONS AND OUTLOOKS}\label{sec:6}

This work presents two black-box machine learning strategies for the control of a wind turbine. These have been benchmarked against a classic PI controller in a numerical environment that reproduces the main features of the wind turbine power production and loading. The results show that these methods can outperform baseline (piecewise linearized) approaches and handle multiple objectives with minimal tuning. Because these controllers learn by trial and error and do not rely on (necessarily simplified) wind turbine models, they could offer new avenues for optimal wind energy extraction. \newline
In the future, hybrid approaches will be considered as proposed in \cite{pinoetal}. These could consist of two agents cooperating on two similar problems: (1) learning the control function and (2) identifying and tuning simplified models online. The controller could thus decide to operate in a model-based or a model-free approach and use one of the methods to improve the other. Both the proposed model-free and the hybrid methods will be implemented in a scaled wind turbine model in an experimental campaign in the L1 wind tunnel at the von Karman Institute. \newline 
\section*{Acknowledgements}
This work has received funding from the Flemish Government through the Agency for Innovation and Entrepreneurship (Vlaams Agentschap Innoveren en Ondernemen, VLAIO) through the ICON project RAINBOW in the context of the cluster on Strategic Initiative for Materials in Flanders (SIM) and the Blauwer cluster.

\end{document}